\documentclass[preprint, showpacs,preprintnumbers,amsmath,amssymb,nofootinbib]{revtex4}

 \usepackage{epsf}
 \usepackage{graphicx}
\usepackage{float}
\begin{document}
\newcommand{\be}[1]{\begin{equation}\label{#1}}
 \newcommand{\ee}{\end{equation}}
 \newcommand{\bea}{\begin{eqnarray}}
 \newcommand{\eea}{\end{eqnarray}}
 \def\disp{\displaystyle}

 \def\gsim{ \lower .75ex \hbox{$\sim$} \llap{\raise .27ex \hbox{$>$}} }
 \def\lsim{ \lower .75ex \hbox{$\sim$} \llap{\raise .27ex \hbox{$<$}} }

\title{\Large \bf Testing the (generalized) Chaplygin gas  model with the Lookback time-Redshift data}
\author{Zhengxiang Li$^{1, 2}$, Puxun Wu$^{1, 2}$  and Hongwei Yu$^{1, 2} \footnote{Corresponding author}$ }

\address{$^1$Department of Physics and Institute of  Physics, Hunan
Normal University, Changsha, Hunan 410081, China
\\$^2$Key Laboratory of Low Dimensional Quantum Structures and Quantum
Control of Ministry of Education, Hunan Normal University, Changsha
410081 }

\begin{abstract}
The Chaplygin gas (CG) and the generalized Chaplygin gas (GCG)
models,  proposed as  candidates of the unified dark matter-dark
energy (UDME), are tested with the look-back time (LT) redshift
data. We find that the LT data only give a very weak constraint on
  the model parameter. However, by combing
the LT with the baryonic acoustic oscillation peak,  we obtain, at
the $95.4\%$ confidence level, $0.68\leq A_c\leq0.82$ and $0.59\leq
h\leq0.65$ for the CG model , and $0.67\leq A_s\leq0.89$ and
$-0.29\leq \alpha\leq0.61$ for the GCG model. This shows that both
the CG and the GCG are viable as a candidate of UDME.  Within the
GCG model, we also find that the Chaplygin gas model ($\alpha=1$) is
ruled out by these data at the $99.7\%$ confidence level.
\end{abstract}

\pacs{98.80.-k; 95.36.+x}

 \maketitle
 \renewcommand{\baselinestretch}{1.5}

\section{Introduction}\label{sec1}
The fact that the universe is undergoing an accelerating expansion
has been supported by many astrophysical and cosmological
observations, such as Type Ia Supernovae (Sne Ia)~\cite{Sne} and
cosmic microwave background radiation (CMBR)~\cite{CMBR1, CMBR2},
etc. In order to explain this mysterious phenomenon, one usually
assumes the existence of an exotic energy with negative pressure,
named dark energy, which dominates the total energy density and
causes an accelerating expansion of our universe at late times. So
far many candidates have been proposed for dark energy, such as the
quintessence~\cite{Quint}, phantom~\cite{Cald},
quintom~\cite{Quintom}, holographic~\cite{Holog} and
agegraphic~\cite{Ageg} models, {et al}.

In 2001 Kamenshchik {\it et al.}~\cite{chaplygin} proposed an
interesting model of dark energy, named the Chaplygin gas (CG),  for
which the equation of state has the form
\begin{equation}
p_{cg}=-\frac{A}{\rho_{cg}}\;,
\end{equation}
where A is a positive constant. Inserting the above equation into
the energy conservation equation, one can obtain the following
expression for the CG  energy density
\begin{equation}
\rho_{cg}=\rho_{cg0}\bigg(A_c+\frac{1-A_c}{a^6}\bigg)^{\frac{1}{2}},
\end{equation}
where $\rho_{cg0}$ is the present energy density of the CG and
$A_c\equiv \frac{A}{\rho_{cg0}}$. Because  the energy density
evolution behaves between dark-matter-like ($\rho_{cg}\propto
a^{-3}$) at early times and dark-energy-like ($\rho_{cg}=-p_{cg} $)
at late times, the CG model has been proposed as a model of the
unified dark matter and dark energy (UDME). For $A_s=0$, the CG
behaves always like matter while for $A_s=1$ it behaves always like
a cosmological constant.

Later Bento {\it et al.}~\cite{generalized} generalized the CG model
by adding a new parameter in the equation of state:
\begin{equation}
p_{gcg}=-\frac{A}{\rho_{gcg}^\alpha}\;.
\end{equation}
Here $\alpha$ is also a positive constant and  $\alpha=1$
corresponds to the CG model. This generalized model was called the
generalized Chaplygin gas (GCG). Combining  the equation of state
and the energy conservation equation for the GCG, we have
\begin{equation}
\rho_{gcg}=\rho_{gcg0}\bigg(A_s+\frac{1-A_s}{a^{3(1+\alpha)}}\bigg)^{\frac{1}{1+\alpha}}\;,
\end{equation}
where $\rho_{gcg0}$ is the present energy density of the GCG and
$A_s\equiv A/\rho_{gcg0}^{1+\alpha}$. Again, the density evolution
in this model changes from $\rho_{gcg}\propto a^{-3}$ at early times
to $\rho_{gcg}=\emph{constant}$ at late times. As a result, the GCG
has also been proposed as the UDME.   Apparently, for $\alpha = 0$
the GCG model behaves like  the cold dark matter plus a cosmological
constant,  while for $A_s=0$, it behaves always like matter and for
$A_s=1$, it behaves always like a cosmological constant.

Currently, works have been done on the CG and the GCG
models~\cite{GCG} and many authors have attempted to constrain them
by using various observational data, such as Sne
Ia~\cite{Ia1,Ia2,Ia3,Ia4,Ia5,Ia6}, the CMBR~\cite{Ia6,cmbr1,cmbr2},
the gamma-ray bursts~\cite{gamma}, the gravitational
lensing~\cite{Ia3,Ia5,lensing}, the X-ray gas mass fraction of
clusters~\cite{Ia2,Ia3,Ia4}, the large scale
structure~\cite{Ia6,structure}, the age of high-redshift
objects~\cite{high-redshift}, and the Hubble parameter versus
redshift~\cite{Hubble}.

Recently, based on the age of the old objects and the age of the
universe from the WMAP5~\cite{CMBR2}, the lookback time-redshift
(LT) data have been used  to test different dark energy
models~\cite{LBT}. The advantage of this data set is that the ages
of distant objects are independent of each other, and thus may avoid
biases that are present in techniques that use distances of primary
or secondary indicators in the cosmic distance ladder method. As a
result, these age data are different from the widely used distance
one. Therefore, it may  still be interesting to test the CG and the
GCG models with these LT data although many other datasets have been
used to test them. This is what we are going to do in the present
paper.  In order to break the degeneracy of model parameters, we
also add,  in our discussion, the baryonic acoustic oscillation
(BAO) peak detected by large-scale correlation function of luminous
red galaxies from Sloan Digital Sky Survey (SDSS)~\cite{SDSS}.

\section{The lookback time-redshift data}\label{2}
The lookback time as a function of redshift can be expressed as the
interval from the age at redshift $z$ ($t_z$) to the present age of
the Universe ($t_0$),
\begin{equation}
t_L(z;\textbf{p})=H_0^{-1}\int_0^z\frac{dz'}{(1+z')E(\textbf{p})},
\end{equation}
where $H_0=h/9.78$ $Gry^{-1}$ is the present value of the Hubble
parameter and \emph{h } can be obtained from the {\it HST}
\emph{key} project (at $1\sigma$ interval $0.64\leq
h\leq0.80$~\cite{HST}). In  the above expression, the form of
$E(\textbf{p})$ is dependent on the selected cosmological model
\begin{equation}
E^2(\textbf{p})=\sum_i \Omega_{i,0}a^{-3(1+w_i)},
\end{equation}
where $E(\textbf{p})\equiv H(\textbf{p})/H_0$ and \textbf{p}
represents the  complete set of parameters for a given cosmological
model with the subscript 0 denoting the present-day value.

The age $t$ at redshift $z_i$ of an object is defined as the
difference between the age of the universe at redshift $z_i$ and
that at redshift $z_F$ when the object was born
\begin{equation}
{t(z_i)}=\frac{1}{H_0}\bigg[{\int_{z_i}^\infty\frac{dz'}{(1+z')E(\textbf{p})}-{\int_{z_F}^\infty\frac{dz'}{(1+z')E(\textbf{p})}}}\bigg].
\end{equation}
Using the above expression and Eq.~(5), $t(z_i)$ can also be written
as
\begin{equation}
t(z_i)=t_L(z_F)-t_L(z_i).
\end{equation}
The observed lookback time to an object at redshift $z_i$ can then
be defined from the above relation~\cite{obl}
\begin{eqnarray}
{t^{obs}_L(z_i)}&=&t_L(z_F)-t_L(z_i)\nonumber\\
&=&[t_0^{obs}-t(z_i)]-[t_0^{obs}-t_L(z_F)]\nonumber\\
&=&t_0^{obs}-t(z_i)-\tau,
\end{eqnarray}
where $\tau$ means the time from Big Bang to the formation of the
objcet, which is the so-called delay factor or incubation time, and
$t_0^{obs}=13.7\pm 0.2 $ Gyr~\cite{Dunkley}.

Using the differential age method, recently, Simon et
al.~\cite{Simon2005} obtained 32 age data points of the passively
evolving galaxies in the redshift interval $0.117<z<1.845$, which
are shown in Fig.~(1). In order to constrain the model parameters
using these age data, we define the likelihood function
 \be{eq12}
 \mathcal{L}_{age}\propto\exp[-\chi_{age}^2(z; \textbf{p},\tau)/2],
 \ee
where $\chi_{age}^2$ is relative to the LT sample:
 \begin{equation}
\chi_{age}^2=\sum\limits_i
\frac{\big[t_L(z_i;\textbf{p})-t_L^{obs}(z_i;\tau)\big]^2}{\sigma_{T}^2}+\frac{\big[t_0(\textbf{p})-t_0^{obs}\big]}{\sigma_{t_o^{obs}}^2}.
 \end{equation}
 In the above expression, $\sigma_T^2\equiv\sigma_i^2+\sigma_{t_0^{obs}}^2$,
 $\sigma_i$ is the uncertainty in the individual lookback time to
 the $i^{th}$ galaxy of our sample and $\sigma_{t_0^{obs}}$ stands
 for the uncertainty on the total expansion age of the
 universe($t_0^{obs}$).
Note that while the observed lookback time ($t_L^{obs}(z_i;\tau)$)
is directly dependent on $\tau$, its theoretical value ($t_L(z_i;
\textbf{p})$) is not. Furthermore, in principle, it must be
different for each object in the sample. Thus the delay factor
becomes  a ``nuisance" parameter,  we use the following method to
marginalize over it~\cite{modlike}
 \begin{eqnarray}
\tilde{\chi}^2&=&-2\ln\int_0^\infty
d\tau\exp\bigg(-\frac{1}{2}\chi_{age}^2\bigg)\nonumber\\
&=&A-\frac{B^2}{C}+D-2\ln\bigg[\sqrt{\frac{\pi}{2C}}
\textrm{erfc}\bigg(\frac{B}{\sqrt{2C}}\bigg)\bigg],
 \end{eqnarray}
where
 \be{eq8}
A=\sum_{i=1}^n
\frac{\triangle^2}{\sigma_{T}^2}.~~~~~~~~~~~~~~B=\sum_{i=1}^n\frac{\triangle}{\sigma_{T}^2}.~~~~~~~~~~~~~~C=\sum_{i=1}^n\frac{1}{\sigma_{T}^2},
 \ee
 D is the second term of the rhs of Eq.(11),
 \begin{equation}
\triangle=t_L(z_i;\textbf{p})-[t_0^{obs}-t(z_i)],
 \end{equation}
and erfc(x) is the complementary error function of the variable x.

\section{Constraint on the CG model}\label{sec2.2}
\subsection{Constraint from lookback time-redshift test}\label{2.1}
For the CG model, the Friedmann equation for a spatially flat
universe which contains only the baryonic matter and the Chaplygin
gas, can be expressed as
 \be{eq16}
H^2(H_0,A_c,h,z)=H_0^2E^2(A_c,z),
 \ee
with $E(A_c,z)$ being given by
 \begin{eqnarray}
 E(A_c,z)=\big[\Omega_b(1+z)^3+(1-\Omega_b) \big(A_c+(1-A_c)(1+z)^6)^{\frac{1}{2}}\big]^{1/2},
 \end{eqnarray}
where $\Omega_b$ is the present dimensionless density parameter of
the baryonic matter and the WMAP observations give $\Omega_b h^2 =
0.0233 \pm 0.0008$~\cite{CMBR2}. For LT data only, Fig.~(2) shows
the contour plot (68.3, 95.4 and $99.7\%$ confidence level) in the
$A_c-h$ plane for the $\chi_{age}^2$ given by Eqs. (10)-(12) with a
Gaussian prior on the Hubble parameter. The best-fit value is
$A_c=0.93$ and $h=0.74$ which agrees very well with the recent
analysis of an extensive new program with the Hubble Space
Telescope~\cite{Riess09}. We obtain $0.32\leq A_c\leq0.98$ and
$0.51\leq h\leq0.88$ at the $95.4\%$ confidence level. Apparently,
LT only gives a very weak constraint on the model parameter $A_c$

\subsection{Joint analysis with BAO}\label{sec2.2}
In order to obtain a tighter constraint on the model parameter, we
combine the LT data with the BAO data.  For the BAO data, the
parameter $\mathcal{A}$ is used, which is independent of
cosmological models and for a flat Universe it can be expressed as
\begin{equation}
\mathcal{A}=\frac{\sqrt{\Omega_m}}{E(z_1)^{1/3}}\bigg[\frac{1}{z_1}\int_0^{z_1}\frac{dz}{E(z)}\bigg]^{2/3}\;,
\end{equation}
where $z_1=0.35$ and $\mathcal{A}$ is measured to be
$\mathcal{A}=0.469\pm0.017$ from the SDSS. In the above expression,
$\Omega_m$ is the effective matter density and it is given by
$\Omega_m=\Omega_b+(1-\Omega_b)(1-A_c)^{1/2}$~\cite{Ia3,Ia4,omb} for
the CG model as a UDME .

In Fig.~(3), we show the results by combining the LT and BAO, which
indicate that both $A_c$ and \emph{h} are tightly constrained. At
$95.4\%$ confidence level, we obtain $0.68\leq A_c\leq0.82$ and
$0.59\leq h\leq0.65$ with the best value given by $A_c=0.74$ and
$h=0.61$.

\section{Constraint on the GCG model}\label{sec2}
For the GCG model,   $E(\textbf{p})$ in the Friedmann equation now
has the form
 \begin{eqnarray}
 E(A_s,\alpha,z)=&&\big[\Omega_b(1+z)^3+(1-\Omega_b)\nonumber\\
&&\times\big(A_s+(1-A_s)(1+z)^{3(1+\alpha)})^{\frac{1}{1+\alpha}}\big]^{1/2}\;.
 \end{eqnarray}
Since we are interested in the parameters $A_s$ and $\alpha$, $h$
becomes a nuisance parameter and is fixed at $h=0.72$ in our
discussion.

In Fig.~(4), we show the results from the lookback time-redshift
(LT) data, which is very similar to the results obtained from the
Sne Ia data~\cite{Ia1}. The best-fit values are $A_s=0.47$ and
$\alpha=-0.85$. From this figure one can see that two model
parameters $A_s$ and $\alpha$ are degenerate and at the $68.3\%$
confidence level, the scenario of the standard dark energy
(cosmological constant) plus dark matter scenario ($\alpha=0$) is
excluded.

The constraints from only  the BAO data on the parameter space
$A_s-\alpha$ shows that although it constrains efficiently the
parameter plane into a narrow strip, parameters $A_s$ and $\alpha$
are also degenerate~\cite{Hubble}. However, it is interesting to see
that possible degeneracies between the model parameters may be
broken by combining these two sets of observational data (LT+BAO).
Fig.~(5)  shows the results in the $A_s-\alpha$ plane for the joint
LT+BAO analysis. The best fit happens at $A_s=0.77$ and
$\alpha=0.02$. At the $95.4\%$ confidence level, we obtain  a
stringent constraint on the GCG: $0.67\leq A_s \leq0.89$ and
$-0.29\leq\alpha\leq0.61$. Apparently,
 the  Chaplygin gas model ($\alpha=1$) is ruled out at the
$99.7\%$ confidence level. In addition, we find  that, at the
$68.3\%$ confidence level,  the combination of these two databases
allows the scenario of the cosmological constant plus dark matter
($\alpha=0$), although the lookback time-redshift alone excludes it.

 \section{Conclusions}
In this paper, we have examined,  with the look-back time redshift
data, the constraints on the Chaplygin gas (CG) model and the
generalized Chaplygin gas (GCG) model, which are proposed as
candidates of the unified dark matter-dark energy (UDME). We find
that, although the current lookback time-redshift data can not
tightly constrain  the CG and the GCG model parameters, stringent
constraints can be obtained by combining the lookback time-redshift
data with baryonic acoustic oscillation peak data. For the CG and
the GCG models, at the $95.4\%$ confidence level, we obtained
$0.68\leq A_c\leq0.82$ and $0.59\leq h\leq0.65$, and $0.67\leq
A_s\leq0.89$ and $-0.29\leq \alpha\leq0.61$, respectively. This
shows that both the CG and the GCG  are viable as  candidates of
UDME. However, from the results obtained within the GCG, we find
that the Chaplygin gas model ($\alpha=1$) is ruled out by these data
at the $99.7\%$ confidence level.  These results are consistent with
those obtained from the X-ray gas mass fractions of galaxy clusters
and the dimensionless coordinate distance of Sne Ia and FRIIb radio
galaxies~\cite{Ia4}, the CMBR power spectrum measurements from
BOOMERANG and Archeops and the Sne Ia~\cite{cmbr2}, and the Hubble
parameter data and the Sne Ia~\cite{Hubble}.

\section*{Acknowledgments}
This work was supported in part by the National Natural Science
Foundation of China under Grants No. 10575035, 10775050, 10705055,
the SRFDP under Grant No. 20070542002,  the Programme for the Key
Discipline in Hunan Province, the FANEDD under Grant No 200922 and
the Research Fund of Hunan Provincial Education Department.

\begin{center}
 \begin{figure}[htbp]
 \centering
 \includegraphics[width=0.45\textwidth]{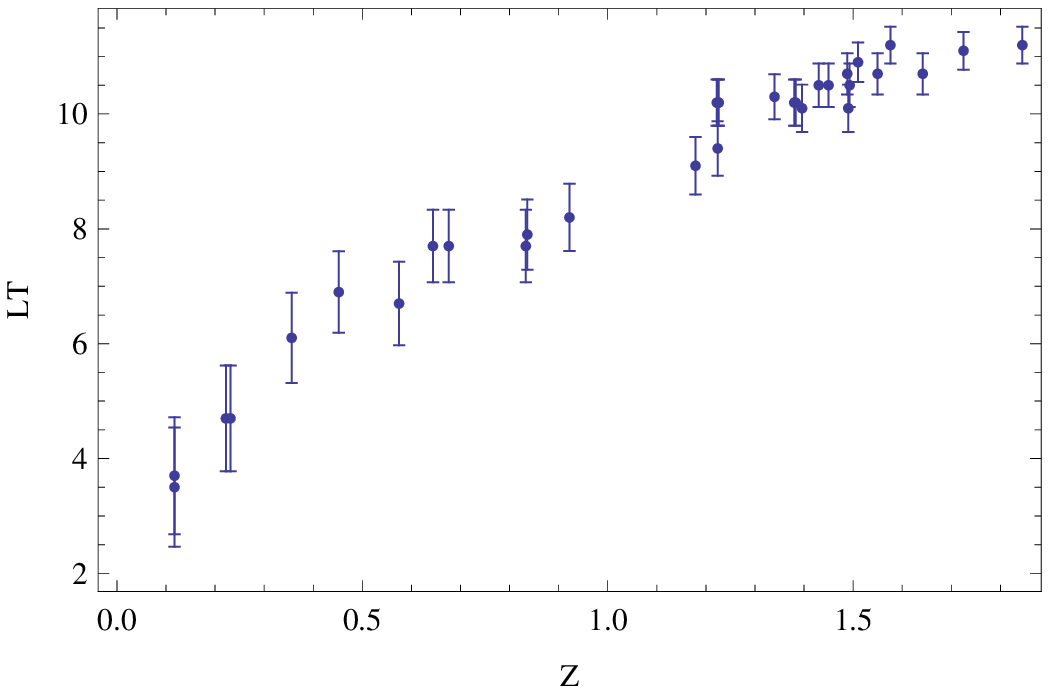}
 \caption{\label{Fig.1.}  The 32 look-back time redshift (LT) data points  obtained from the old passive galaxies
distributed over the redshift interval $0.11\leq z \leq 1.84$  with
the estimate of the total
 age of universe $t_0^{obs}=13.7\pm0.2 Gry$ from the CMB data~\cite{Dunkley}.}
 \end{figure}
 \end{center}

\begin{center}
 \begin{figure}[htbp]
 \centering
 \includegraphics[width=0.45\textwidth]{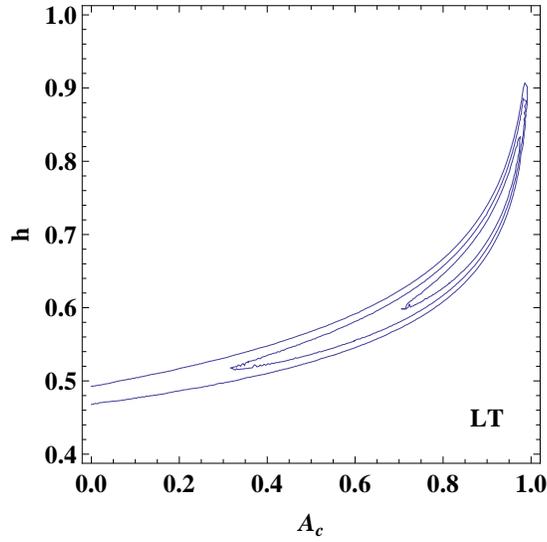}
 \caption{\label{Fig.2.}  The $68.3$, $95.4$ and $99.7\%$ confidence level
  contours for $A_c$ versus \emph{h} from the lookback time-redshift data.
   The best fit happens at $A_c=0.77$ and $h=0.74$.}
 \end{figure}
 \end{center}

\begin{center}
 \begin{figure}[htbp]
 \centering
 \includegraphics[width=0.45\textwidth]{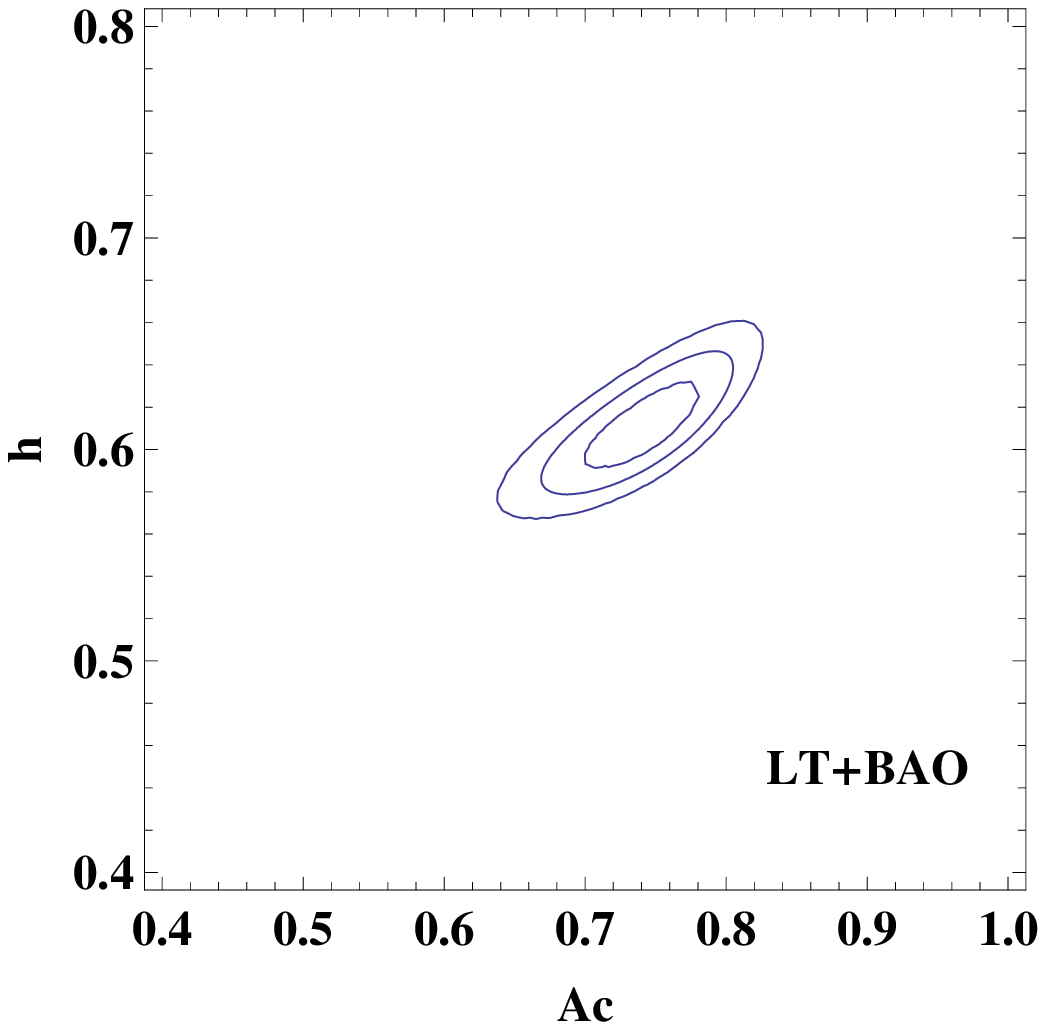}
 \caption{\label{Fig.3.}  The $68.3$, $95.4$ and $99.7\%$ confidence level
  contours for $A_c$ versus \emph{h} from the lookback time-redshift data plus SDSS baryonic acoustic
  oscillations peak.
   The best fit happens at $A_c=0.74$ and $h=0.61$.}
 \end{figure}
 \end{center}

 \begin{center}
 \begin{figure}[htbp]
 \centering
 \includegraphics[width=0.45\textwidth]{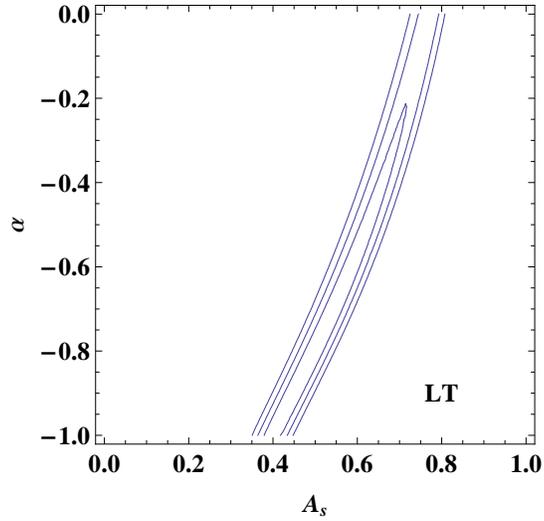}
 \caption{\label{Fig.4.}  The $68.3$, $95.4$ and $99.7\%$ confidence level
  contours for $A_s$ versus $\alpha$ from the lookback time-redshift data with a fixed h=0.72.
  The best fit happens at $A_s=0.47$ and $\alpha=-0.85$.}
 \end{figure}
 \end{center}

\begin{center}
 \begin{figure}[htbp]
 \centering
 \includegraphics[width=0.45\textwidth]{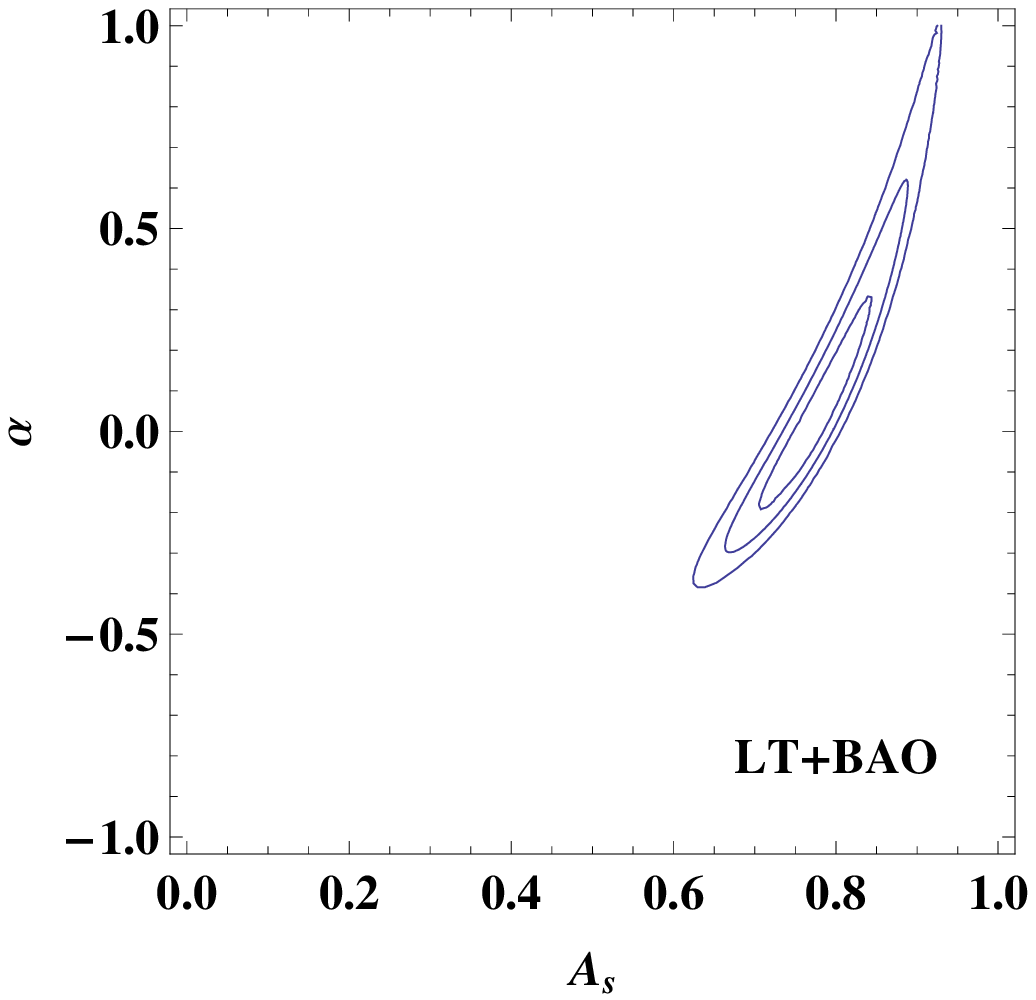}
 \caption{\label{Fig.5.}  The $68.3$, $95.4$ and $99.7\%$ confidence level
  contours for $A_s$ versus $\alpha$ from the lookback time-redshift data
  plus SDSS baryonic acoustic
  oscillations peak.
  The best fit happens at $A_s=0.77$ and $\alpha=0.02$.}
 \end{figure}
 \end{center}

\end{document}